\documentclass[twocolumn]{aastex62}

\usepackage[shortcuts]{extdash}
\usepackage{ctable}

\def\apjl{ApJL}

\def\aap{A\&A}

\def\apjs{ApJS}

\newcommand{\chandra}{\textit{Chandra}}
\newcommand{\xmm}{\textit{XMM-Newton}}
\newcommand{\swift}{\textit{Swift}}
\newcommand{\nustar}{\textit{NuSTAR}}
\newcommand{\nicer}{\textit{NICER}}

\newcommand{\fluxcgs}{erg~s$^{-1}$~cm$^{-2}$}
\newcommand{\lumcgs}{erg~s$^{-1}$}

\received{}
\revised{}
\accepted{}
\submitjournal{ApJ} 

\shorttitle{NGC 4190 ULX-1}
\shortauthors{Earnshaw et al.}

\begin{document}

\title{Return to the forgotten ULX: a broadband \nicer+\nustar\ study of NGC 4190 ULX-1}

\correspondingauthor{Hannah P. Earnshaw}
\email{hpearn@caltech.edu}

\author[0000-0001-5857-5622]{Hannah P. Earnshaw}
\affil{Cahill Center for Astronomy and Astrophysics, California Institute of Technology, Pasadena, CA 91125, USA}

\author[0000-0002-4576-9337]{Matteo Bachetti}
\affil{NAF-Osservatorio Astronomico di Cagliari, Selargius (CA), Italy}

\author[0000-0002-8147-2602]{Murray Brightman}
\affil{Cahill Center for Astronomy and Astrophysics, California Institute of Technology, Pasadena, CA 91125, USA}

\author[0000-0003-0388-0560]{Felix F\"{u}rst}
\affil{European Space Astronomy Centre (ESAC), Science Operations Department, 28692, Villanueva de la Cañada, Madrid, Spain}

\author{Fiona A. Harrison}
\affil{Cahill Center for Astronomy and Astrophysics, California Institute of Technology, Pasadena, CA 91125, USA}

\author{Matthew Middleton}
\affil{School of Physics \& Astronomy, University of Southampton, Southampton, Southampton SO17 1BJ, UK}

\author[0000-0002-8961-939X]{Renee Ludlam}
\affil{Department of Physics \& Astronomy, Wayne State University, 666 West Hancock Street, Detroit, MI 48201, USA}

\author[0000-0002-8403-0041]{Sean N. Pike}
\affil{Department of Astronomy \& Astrophysics, University of California, San Diego, 9500 Gilman Drive, La Jolla, CA, 92093, USA}

\author[0000-0003-2686-9241]{Daniel Stern} 
\affil{Jet Propulsion Laboratory, California Institute of Technology, 4800 Oak Grove Drive, Pasadena, CA 91109, USA}

\author[0000-0001-5819-3552]{Dominic J. Walton}
\affil{Centre for Astrophysics Research, University of Hertfordshire, College Lane, Hatfield AL10 9AB, UK}

\begin{abstract}

We observed the nearby and relatively understudied ultraluminous X-ray source (ULX) NGC 4190 ULX-1 jointly with \nicer\ and \nustar\ to investigate its broadband spectrum, timing properties, and spectral variation over time. We found NGC 4190 ULX-1 to have a hard spectrum characterized by two thermal components (with temperatures $\sim$0.25\,keV and $\sim$1.6\,keV) and a high-energy excess typical of the ULX population, although the spectrum turns over at an unusually low energy. While no pulsations were detected (with pulsed fraction 3$\sigma$ upper limits of 16\% for \nicer\ and 35\% for \nustar), the source shows significant stochastic variability and the covariance spectrum indicates the presence of a high-energy cut-off power-law component, potentially indicative of an accretion column. Additionally, when fitting archival \xmm\ data with a similar model, we find that the luminosity-temperature evolution of the hot thermal component follows the behavior of a super-Eddington slim disk though the expected spectral broadening for such a disk is not seen, suggesting that the inner accretion disk may be truncated by a magnetic field. Therefore, despite the lack of detected pulsations, there is tantalizing evidence for NGC 4190 ULX-1 being a candidate neutron star accretor, although further broadband observations will be required to confirm this behavior. 

\end{abstract}


\section{Introduction} \label{sec:intro}

It is now widely agreed that ultraluminous X-ray sources (ULXs; defined as extragalactic non-nuclear point sources with X-ray luminosity $>$10$^{39}$\,\lumcgs) are a population primarily made up of stellar-mass compact objects accreting at super-Eddington rates, giving them their distinctive high luminosities and particular spectral shapes (for recent reviews, see \citealt{king23,pinto23}). Key to this discovery initially were high-quality \xmm\ observations and, subsequently, \nustar\ observations, which showed that the spectra of ULXs turn over around $\sim$5\,keV (e.g. \citealt{stobbart06,sutton13,bachetti13,walton14}), demonstrating an `ultraluminous' spectral state \citep{gladstone09} distinct from the canonical sub-Eddington accretion states, which exhibit power-law emission up to far higher photon energies \citep{remillard06,done07}.

The ultraluminous spectral state tends to exhibit two thermal components \citep{middleton15}, as well as a steep power-law-like excess above $\sim$10\,keV (e.g. \citealt{walton18b}). Typically, the hard thermal component can be fitted with a broadened disk model, and may correspond to emission from the central region of a super-Eddington slim disk in which advection plays a significant role, which has a broader temperature profile compared to a sub-Eddington thin disk \citep{abramowicz88,watarai00}. The soft thermal component is typically fitted with a standard disk blackbody model, though it likely originates from a massive outflowing wind driven by the radiation pressure \citep{poutanen07,urquhart16}. The presence of such winds was confirmed by the detection of blueshifted absorption features in multiple ULXs, indicating wind velocities of $\sim$0.2$c$ \citep{middleton14,middleton15b,pinto16,pinto17,kosec18}. 

The detection of pulsations in a small number of ULXs \citep{bachetti14,fuerst16,israel17,carpano18,rodriguezcastillo19,sathyaprakash19} confirmed super-Eddington accretion in at least some of the ULX population, since pulsations indicate the presence of a neutron star accretor, with the high luminosities indicating accretion happening at tens to hundreds of times the Eddington limit. The pulsed fraction of neutron star ULXs is found to increase with energy, with the pulsed spectrum having the cut-off power-law shape expected of an accretion column \citep{brightman16,walton18a}. This feature accounts for the steep power-law excess above 10\,keV in these sources, though this excess is also seen in sources from which pulsations have not yet been detected \citep{walton18b}. This may indicate that more of the ULX population have neutron star accretors than just those for which we detect pulsations, or that magnetic and non-magnetic sources are capable of producing similar spectra, and further details are required to distinguish them. Recent simulation work has demonstrated that a hard excess over a Wien tail can be produced by the inner regions of a super-Eddington accretion flow even in the absence of an accretion column (e.g. \citealt{mills23}).

In addition to pulsations, a number of other timing properties are observed in the ULX population. While much of the population of ULXs do not show strong short-term variability (e.g. \citealt{feng05}), some do show significant stochastic variability rising to low frequencies (e.g. \citealt{heil09}). For some sources with sufficiently long observational coverage, there is a break at low frequencies (1--100\,mHz) giving rise to flat-topped noise (e.g. \citealt{earnshaw16}), which is also seen in the Galactic supercritical source SS433 \citep{atapin15}. With some exceptions, the presence of strong variability appears to mainly correspond to sources with softer spectra, which has been proposed to be due to variability being imprinted on the hard central emission by a clumpy outflowing wind along the line of sight (e.g. \citealt{sutton13,middleton11,middleton15}). A small number of ULXs also show quasi-periodic oscillations (QPOs; e.g. \citealt{feng07,rao10,agrawal15}), which may have the potential to provide insight into the geometry of the system.

The \textit{Neutron Star Interior Composition Explorer} (\nicer) telescope mounted on the International Space Station, with its large effective area and 100\,ns timing resolution, is an excellent instrument for spectral and timing studies of X-ray binaries. However, since it is a non-focusing instrument, its application to ULXs is limited due to the lower flux of extragalactic sources and the source confusion caused by multiple bright sources in a single galaxy. Nevertheless, it has successfully been used for the analysis of NGC~300~ULX-1 \citep{ray19,ng22}, which demonstrates its capacity for investigating bright and isolated ULXs. 

NGC 4190 ULX-1 is a bright ULX in the nearby irregular galaxy NGC 4190 at $\sim$2.9\,Mpc \citep{tully16}, located at 12$^h$\,13$^m$\,45$^s$.2 +36$^\circ$\,37$^\prime$\,54$^{\prime\prime}$ \citep{evans19}. The source is persistently bright but variable over time, and has previously exhibited interesting spectral behaviour, with an unusually low turnover energy in the \xmm\ band of 2--4\,keV \citep{ghosh21}, potentially making its high-energy excess particularly accessible for study. It also demonstrates varying high-energy emission above the turnover, in contrast to some other ULXs for which the high-energy emission remains remarkably consistent despite significant variability at lower energies (e.g. \citealt{walton17,walton20}).

\begin{figure*}
	\begin{center}
	\includegraphics[width=14cm]{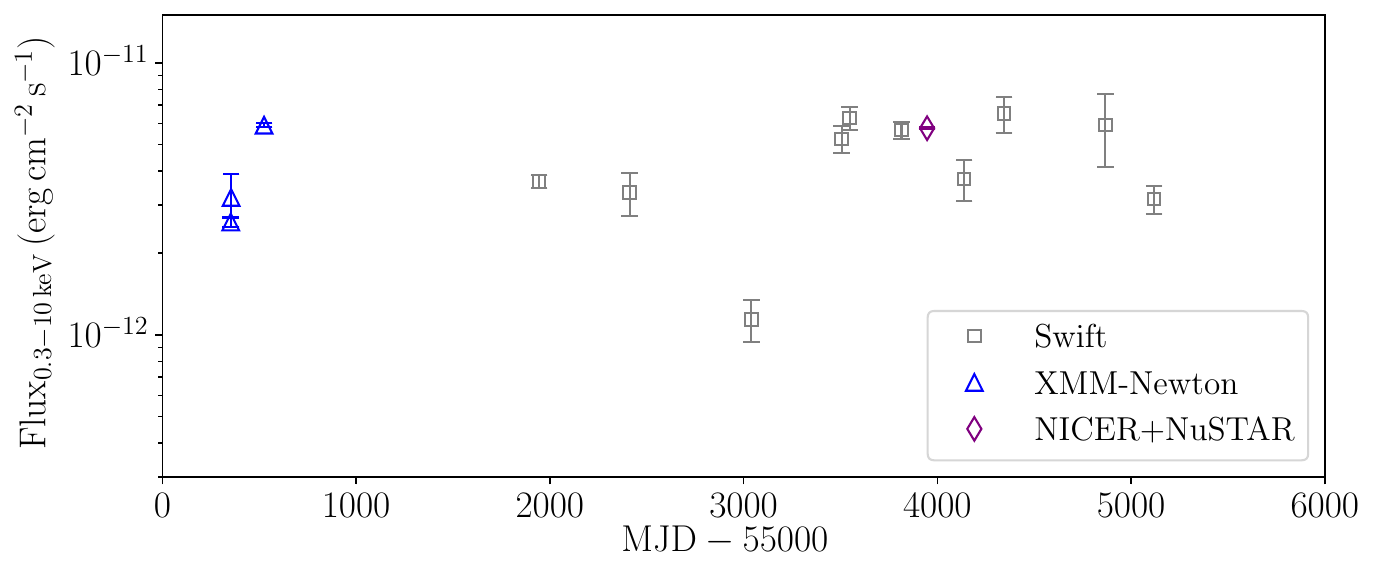}
	\end{center}
	\vspace{-5mm}
	\caption{Long-term light curve of NGC~4190~ULX-1 in the 0.3--10\,keV energy range. \swift-XRT observations are plotted with grey squares, \xmm\ data with blue triangles, \nicer+\nustar\ data with purple diamonds. \swift\ fluxes are retrieved from the products generation tools provided on the \swift\ website; \xmm\ and \nicer+\nustar\ fluxes from spectral fitting and converted to the 0.3--10\,keV energy range where necessary (see Section~\ref{sec:results}). \label{fig:longtermlc}}
\end{figure*}

While previously identified as a high-flux ULX (3--$7\times10^{-12}$\,\fluxcgs) from \xmm\ observations \citep{earnshaw19}, NGC~4190~ULX-1 was initially not followed up with instruments such as \nustar\ due to its proximity to the center of its host galaxy ($\sim$10$^{\prime\prime}$), which in many cases means that the X-ray emission of a source will suffer from confusion from the galactic nucleus for instruments without the resolving power of \chandra. However, examination of an archival observation of NGC~4190 using the High Resolution Camera on \chandra\ shows that NGC~4190~ULX-1 is the only bright X-ray source in the galaxy and within the 5\,arcminute field of view of \nicer\ (one other, far fainter source exists within 25\,arcsec, too faint to contaminate a \nicer\ spectrum). The existing \swift-XRT observations of the source (retrieved from the online light curve generation facility on the \swift\ website; \citealt{evans09}) show it to demonstrate significant long-term variability (Fig.~\ref{fig:longtermlc}).

Its isolation on the sky, combined with a relatively high flux from its proximity to us, makes NGC~4190~ULX-1 an ideal ULX to be observed with \nicer\ and \nustar\ in order to investigate its spectral and timing properties. 

In this paper we detail the analysis of a new simultaneous observation of NGC~4190~ULX-1 with \nicer\ and \nustar, along with a reanalysis of archival \xmm\ data in light of our discoveries. In Section~\ref{sec:data} we describe the observations and data reduction, in Section~\ref{sec:results} we describe the spectral and timing analysis, and in Section~\ref{sec:disc} we present our discussion and conclusions. We note that an independent paper on this source \citep{combi24} was released while this work was under submission, and we provide some comparisons between these two works.

\section{Data Reduction} \label{sec:data}

On 26 April 2020, we performed a simultaneous observation of NGC~4190 with \nicer\ (Observation IDs: 3645010101--4) and \nustar\ (Observation ID: 30601009002), for total good exposure times of 24.7\,ks and 85.0\,ks respectively, in order to investigate its broadband spectral and timing properties (PI: Earnshaw). There are also three archival observations of the galaxy taken with \xmm\ in 2010 which we use to investigate its long term spectral variability. We detail the observations used in this investigation in Table~\ref{tab:obs}. 

\begin{table*}
\caption{The observations used in this investigation, consisting of archival \xmm\ observations from 2010, and the \nicer\ and \nustar\ observations taken for this investigation.} \label{tab:obs}
	\begin{center}
		\begin{tabular}{lccc}
			\hline
			Observatory & Observation ID & Start Time & Exposure Time (ks) \\
			\hline
			\xmm & 0654650101 & 2010-06-06 12:08:27 & 6.2/6.2/0.2$^a$ \\ 
			 & 0654650201 & 2010-06-08 11:14:45 & 15.2/16.9/5.9 \\
			 & 0654650301 & 2010-11-25 01:24:51 & 14.4/14.8/8.0 \\
			\hline
			\nicer & 3645010101 & 2020-04-26 19:02:09 & 24.7$^b$ \\ 
			  & 3645010102 & 2020-04-27 01:13:10 & \\
			  & 3645010103 & 2020-04-28 00:26:11 & \\
			  & 3645010104 & 2020-04-28 23:39:51 & \\
			\nustar & 30601009002 &  2020-04-26 19:31:09 & 78.4 + 11.8$^c$ \\ 
			\hline
		\end{tabular}
	\end{center}
	$^a$\xmm\ exposure times are given for the EPIC MOS1/MOS2/pn instruments, after filtering for background flaring.\\
	$^b$\nicer\ exposure time after combining and filtering all four observations. \\
	$^c$\nustar\ exposure time taken in mode 1 and mode 6 respectively. 
\end{table*}

\subsection{NICER}

The \nicer\ data were reduced using {\sc nicerdas} version 2020-04-23\_V007a, with CALDB version 20200722. Calibration and pre-filtering were performed using the {\sc nicerl2} routine, and barycenter correction applied with {\sc barycorr} (with ephemeris DE-405), using the \chandra\ source coordinates \citep{evans19}. The four observations were then merged using {\sc nimpumerge}, and good time intervals (GTIs) created using the following settings: SUN\_ANGLE$>$60, COR\_SAX$>$4.0 and KP$<$5. These GTIs were used to extract events between 0.2 and 12\,keV using {\sc niextract-events}, and then {\sc xselect} was used to extract spectra and light curves. We used the standard response matrix file (RMF) and on-axis ancillary response file (ARF) as provided in the CALDB, and generated the background spectrum using both the {\sc nicer\_bkg\_estimator} tool and the {\sc nibackgen3C50} tool. 

The count rate for NGC~4190~ULX-1 is comparable to the \nicer\ background rate. Plotting the counts spectrum along with the background shows that the background dominates over the data at low and high energies (Fig.~\ref{fig:counts}), though the two methods for background estimation do not entirely agree on the exact energy range for which the source is dominant. Therefore, we take an approximate average of the low- and high-energy bounds and only consider data in the energy range 0.5--4.5\,keV when analyzing this source. Additionally, we used the background generated from the {\sc nicer\_bkg\_estimator} method during analysis as it is more conservative about the contribution from background than {\sc nibackgen3C50}. Using {\sc nicer\_bkg\_estimator} as a basis, we estimate the background count rate at 0.77\,ct\,s$^{-1}$ between 0.5--4.5\,keV (compared to a net source rate of 1.75\,ct\,s$^{-1}$ in the same band). 

\subsection{NuSTAR}

The \nustar\ data were reduced using {\sc nustardas v2.0.0}, with CALDB version 20211020. The data were reduced using the {\sc nupipeline} routine, and data products extracted using {\sc nuproducts} with barycenter correction applied, using a 60$^{\prime\prime}$ radius source extraction region centered on the PSF defined for each of the FPMA and FPMB cameras. Background regions of 90$^{\prime\prime}$ radius located on the same chip were used to extract background products. The normal science mode (i.e. mode 1) contained 78.4\,ks of good time data. A significant portion of the observation was spent in mode 6 (when aspect reconstruction using Camera Head Unit CHU4 is unavailable), so we also used {\sc nusplitsc} to produce event files for the  alternate CHU combinations. Data products were then extracted from these additional event files and combined with the mode 1 products using the HEASoft tools {\sc lcmath} and {\sc addascaspec}. For a more detailed description of the mode 6 data extraction process, see \citet{walton16}. In this way, we were able to utilize 11.8\,ks of additional data. 

\begin{figure}
	\begin{center}
	\includegraphics[width=7.5cm]{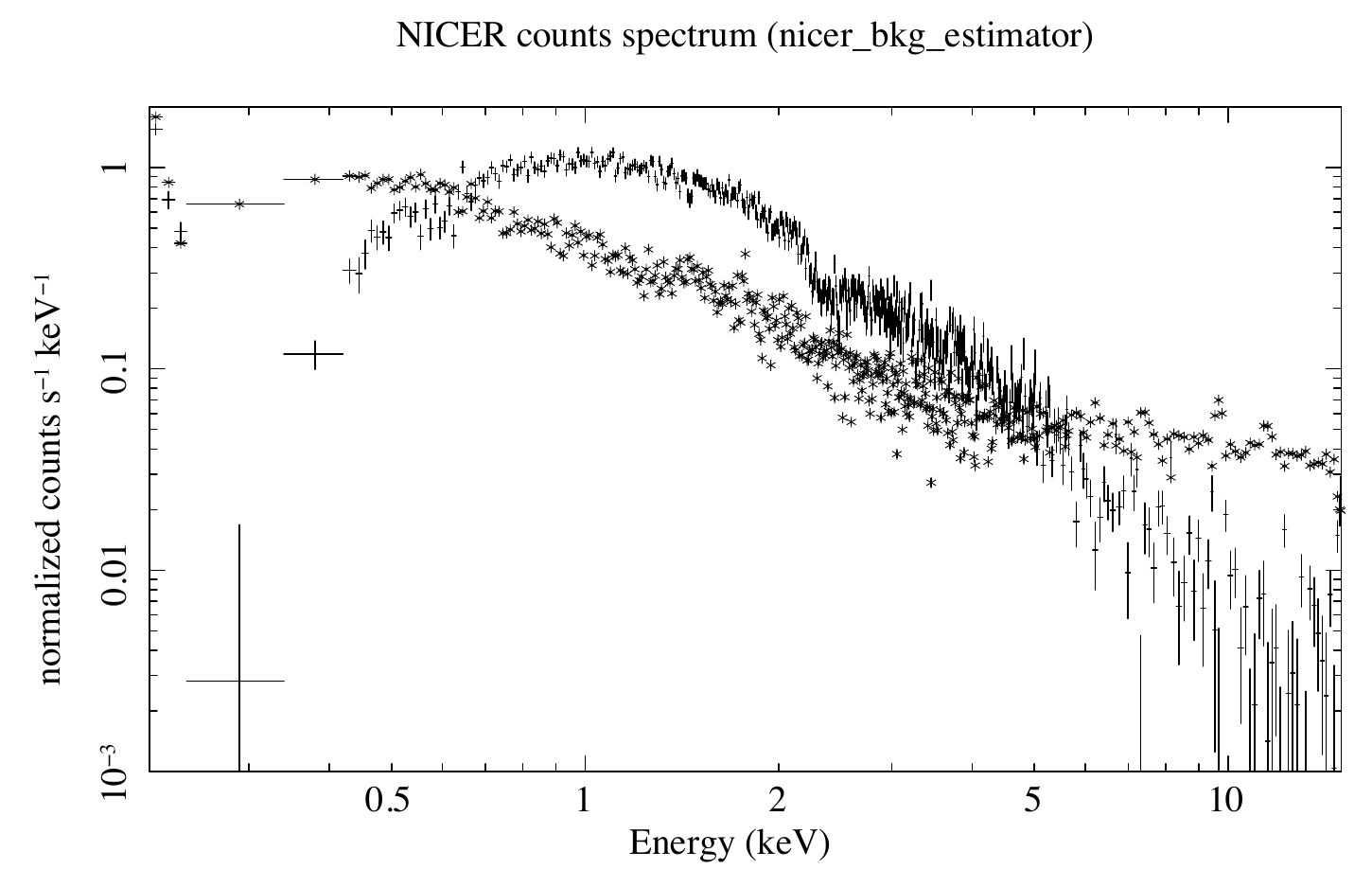}
	\includegraphics[width=7.5cm]{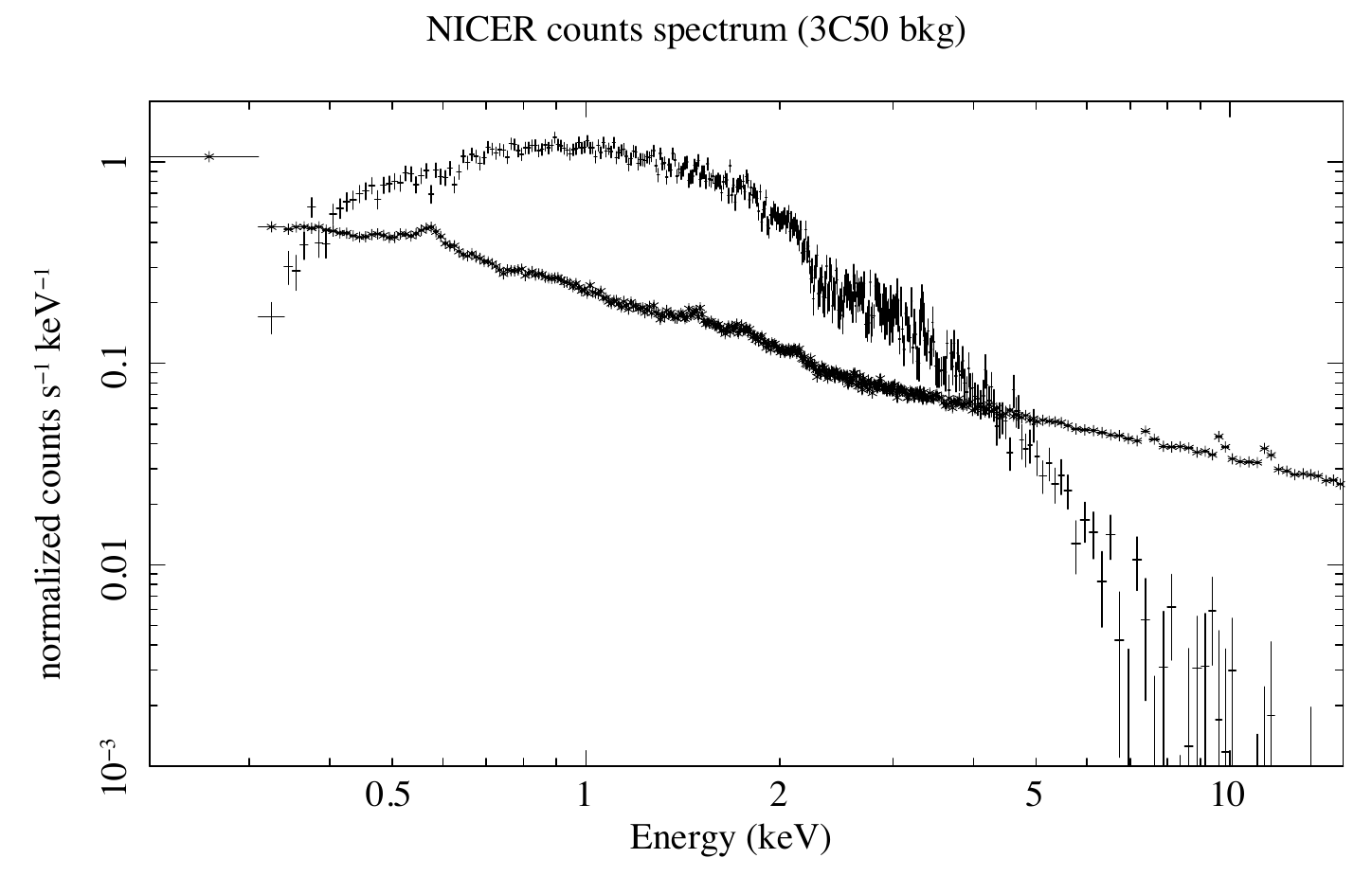}
	\includegraphics[width=7.5cm]{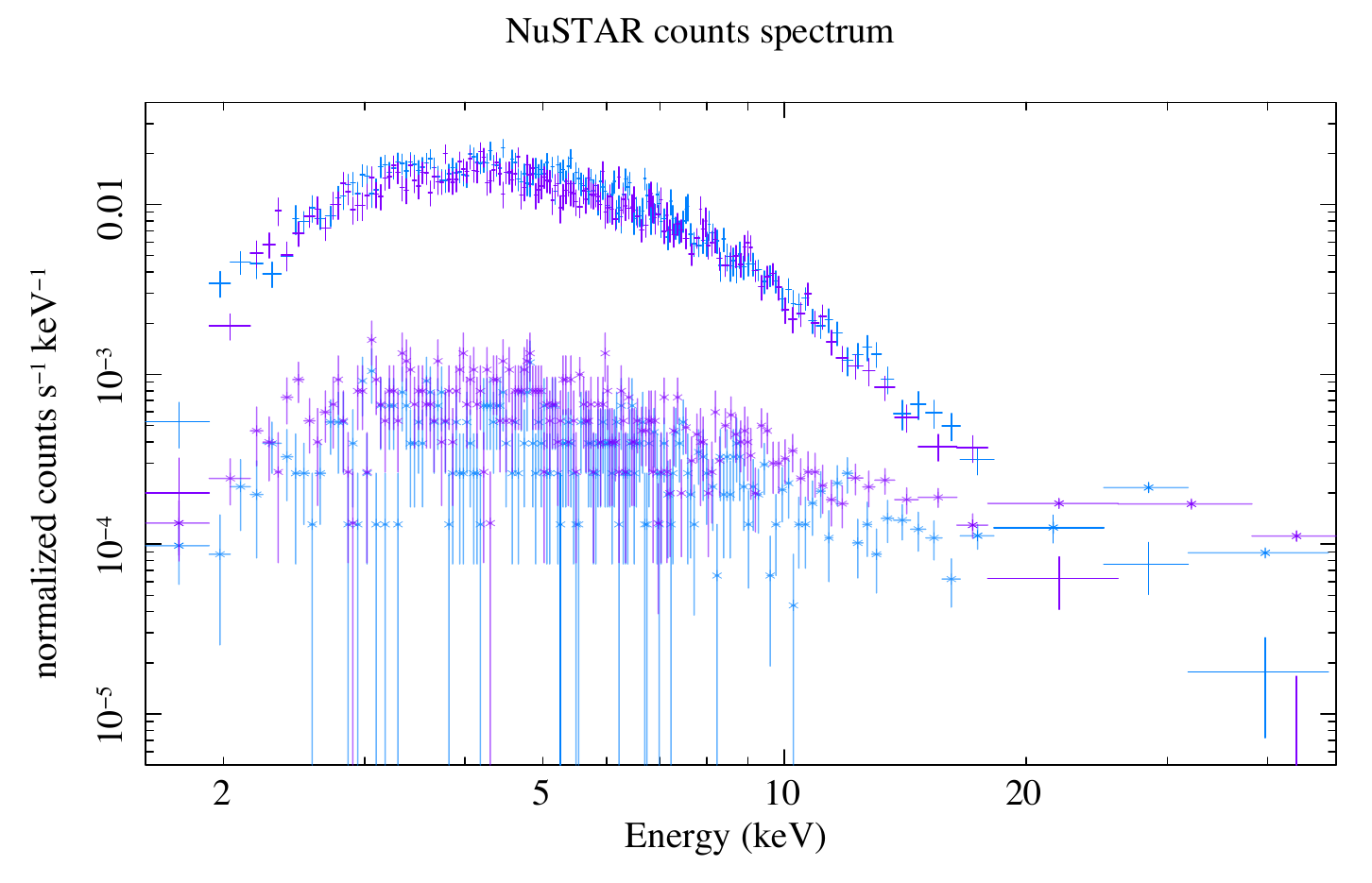}
	\end{center}
	\vspace{-5mm}
	\caption{Counts spectra including background for \nicer\ (black) and \nustar\ FPMA and FPMB (blue and purple). The \nicer\ background is plotted as estimated using {\sc nicer\_bkg\_estimator} (top) and {\sc nibackgen3C50} (middle). \label{fig:counts}}
\end{figure}

Plotting the combined counts spectrum with the background shows the source data dominating over background up to 20\,keV (Fig.~\ref{fig:counts}). Therefore we limited the \nustar\ data to the range 3--20\,keV for this analysis.

\subsection{XMM-Newton}

As well as the \nicer\ and \nustar\ data from our observation, we also used \xmm\ data from three archival observations of NGC~4190~ULX-1. We extracted and reduced EPIC-MOS and pn data using the \xmm\ Science Analysis Software v17.0.0. The data were reduced using the {\sc emproc} and {\sc epproc} routines and, following standard procedures, we filtered for background flaring where the MOS count rate exceeded 0.35\,ct\,s$^{-1}$ in the 10--12\,keV band and where the pn count rate exceeded 0.4\,ct\,s$^{-1}$ in the $>$12\,keV band. Observation 0654650101 was particularly affected by flaring, and we do not use the pn data for this observation. Spectra and light curves in the 0.3--10\,keV energy band were extracted from 40$^{\prime\prime}$ radius source regions, using {\tt FLAG==0} and {\tt PATTERN$<$4} events for the pn camera and {\tt PATTERN$<$12} for the MOS cameras. Background products were extracted from 60$^{\prime\prime}$ regions on the same chips at a similar distance from the readout node. The RMF and ARF were generated using the {\sc rmfgen} and {\sc arfgen} tasks. 

\section{Analysis \& Results} \label{sec:results}

\begin{figure}
	\begin{center}
	\includegraphics[width=8cm]{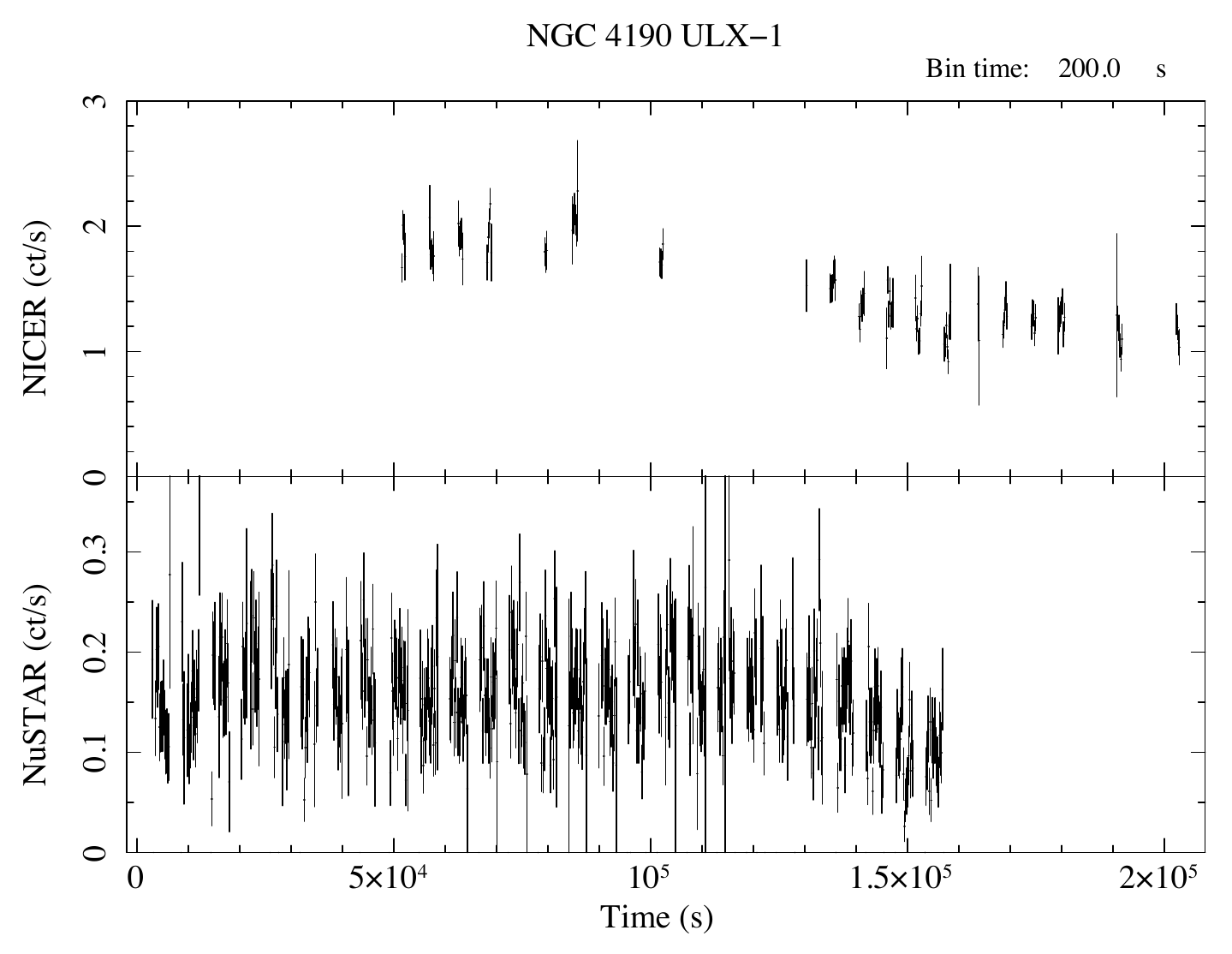}
	\end{center}
	\vspace{-5mm}
	\caption{The light curve for \nicer\ (0.5--4.5\,keV) and \nustar\ (3--20\,keV) over the course of the observation, in 200\,s time bins. Both light curves are background-subtracted (with the background assumed to be a constant 0.77\,ct\,s$^{-1}$ for the \nicer\ light curve).  \label{fig:lc}}
\end{figure}

NGC~4190~ULX-1 is persistently bright over the course of our observation, with an average \nicer\ net count rate of 1.75\,ct\,s$^{-1}$ in the 0.5--4.5\,keV energy range (compared with 0.77\,ct\,s$^{-1}$ due to background), and \nustar\ count rate of 0.3\,ct\,s$^{-1}$. The source declines in flux slightly about halfway through the \nicer\ observation, towards the end of the \nustar\ observation (Fig.~\ref{fig:lc}). 

\subsection{Spectral Analysis}

We performed all spectral fitting using v12.10 of the XSPEC \citep{arnaud96} software, and all quoted models are given in XSPEC syntax. In all cases, spectra were grouped into at least 20 counts per bin to allow for $\chi^2$ statistics to be used in fitting. Uncertainties are given at the 90\% confidence level, and we use the abundance tables of \citet{wilms00} throughout. Where we use two {\tt tbabs} absorption models, the first is always frozen to the Galactic value of $N_{\rm H} = 1.84\times10^{20}$\,cm$^{-2}$ \citep{kalberla05}. We also use a constant {\tt const} model, fixed to 1 for the \nicer\ data, to account for calibration differences between the instruments. We give the parameters of all fits in Table~\ref{tab:spectralfitting}.

We fitted the \nicer\ spectrum simultaneously with the \nustar\ FPMA and FPMB spectra. Since the broadband spectrum exhibits clear curvature, we began by fitting it with single-component absorbed broadened disk and cut-off power-law models, as sometimes observed in high-luminosity ULXs (e.g. \citealt{sutton17,pintore17}), and expected to be produced by a supercritical accretion disk or an accretion column respectively. Neither model provided a good fit to the data ($\chi^2_\nu\sim1.4$), and the residuals show an `m-shaped' structure that indicates spectral curvature not described by the model. 

\begin{figure}
	\begin{center}
	\includegraphics[width=7.9cm]{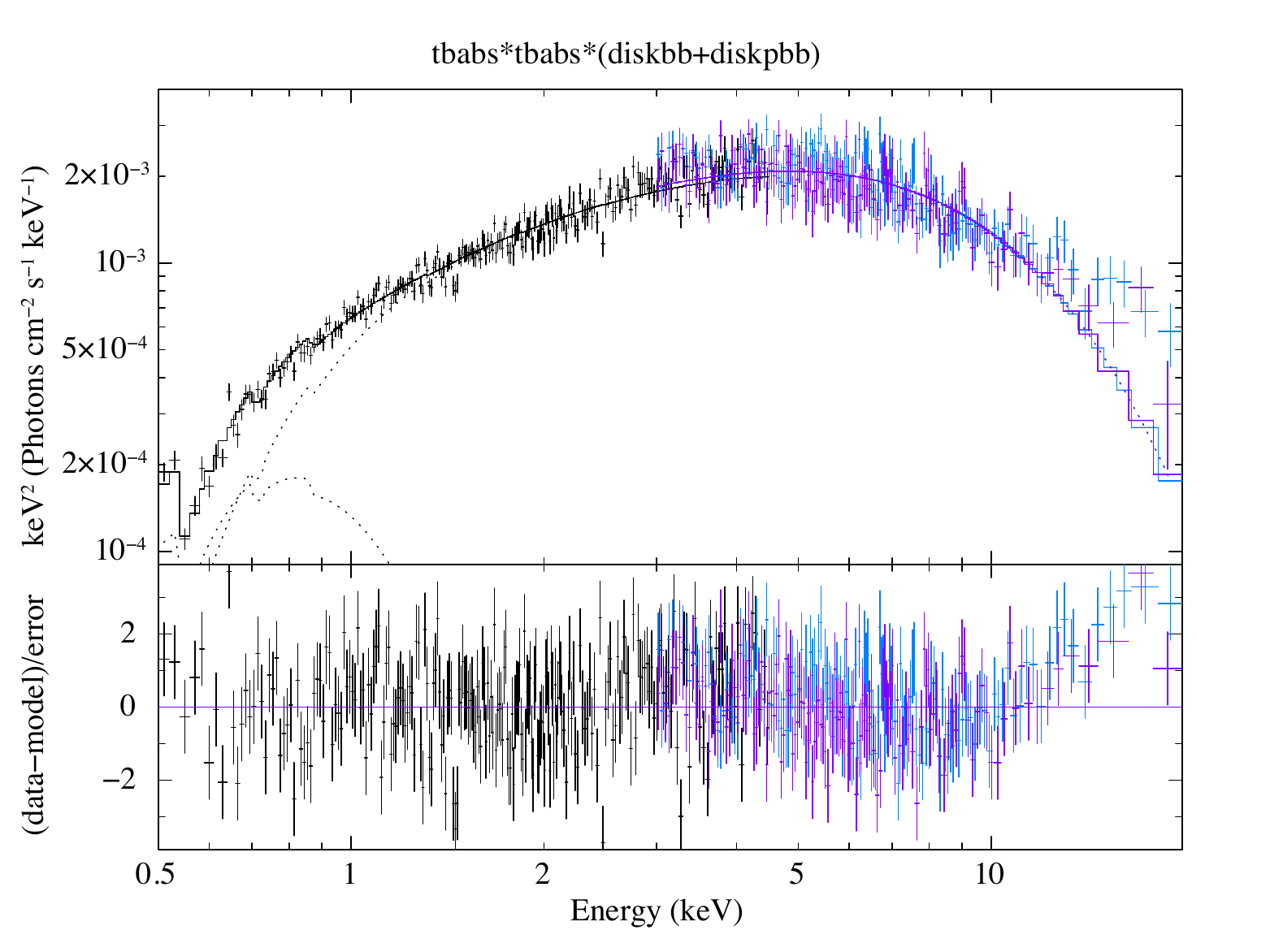}
	\includegraphics[width=7.9cm]{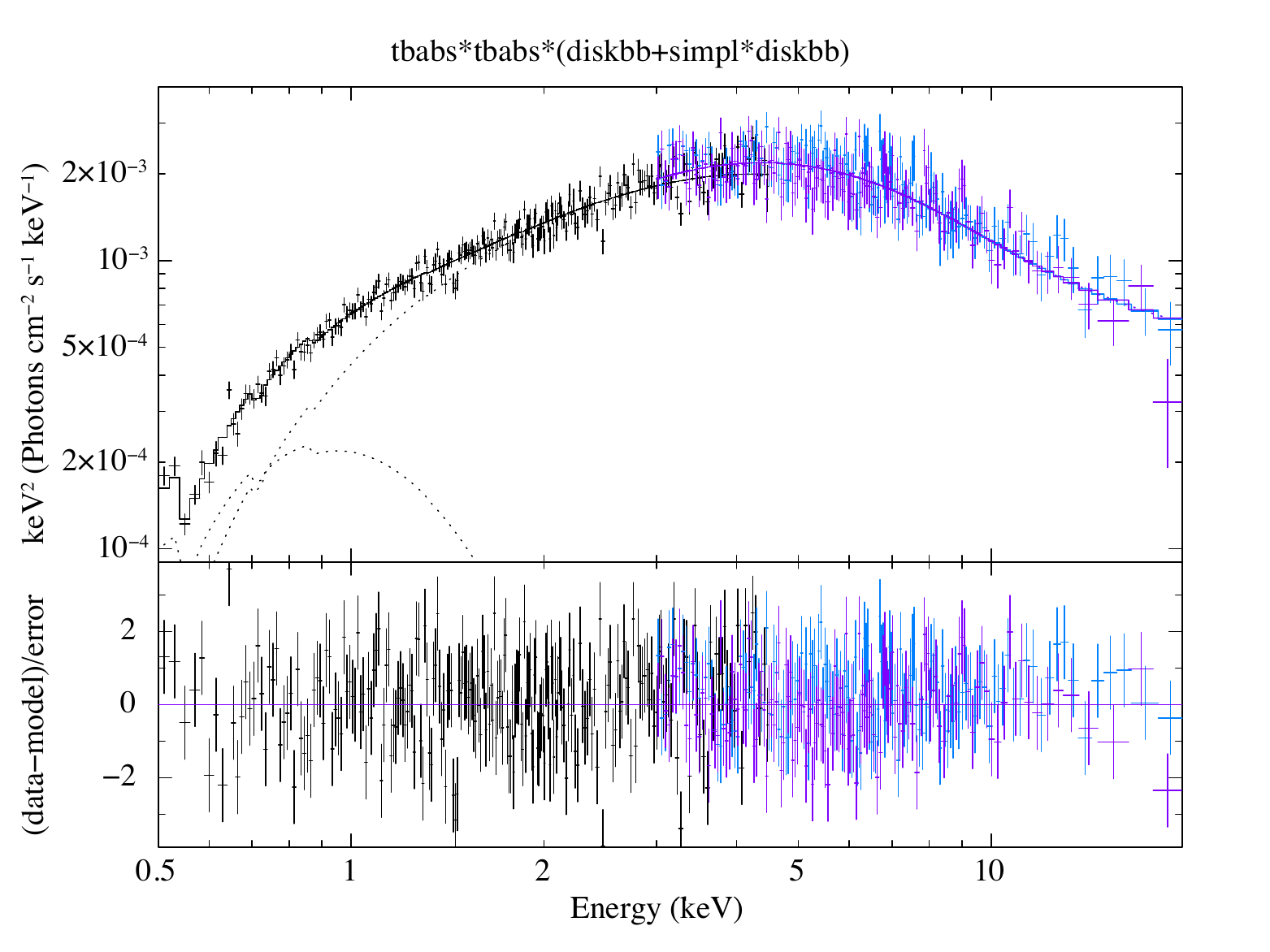}
	\includegraphics[width=7.9cm]{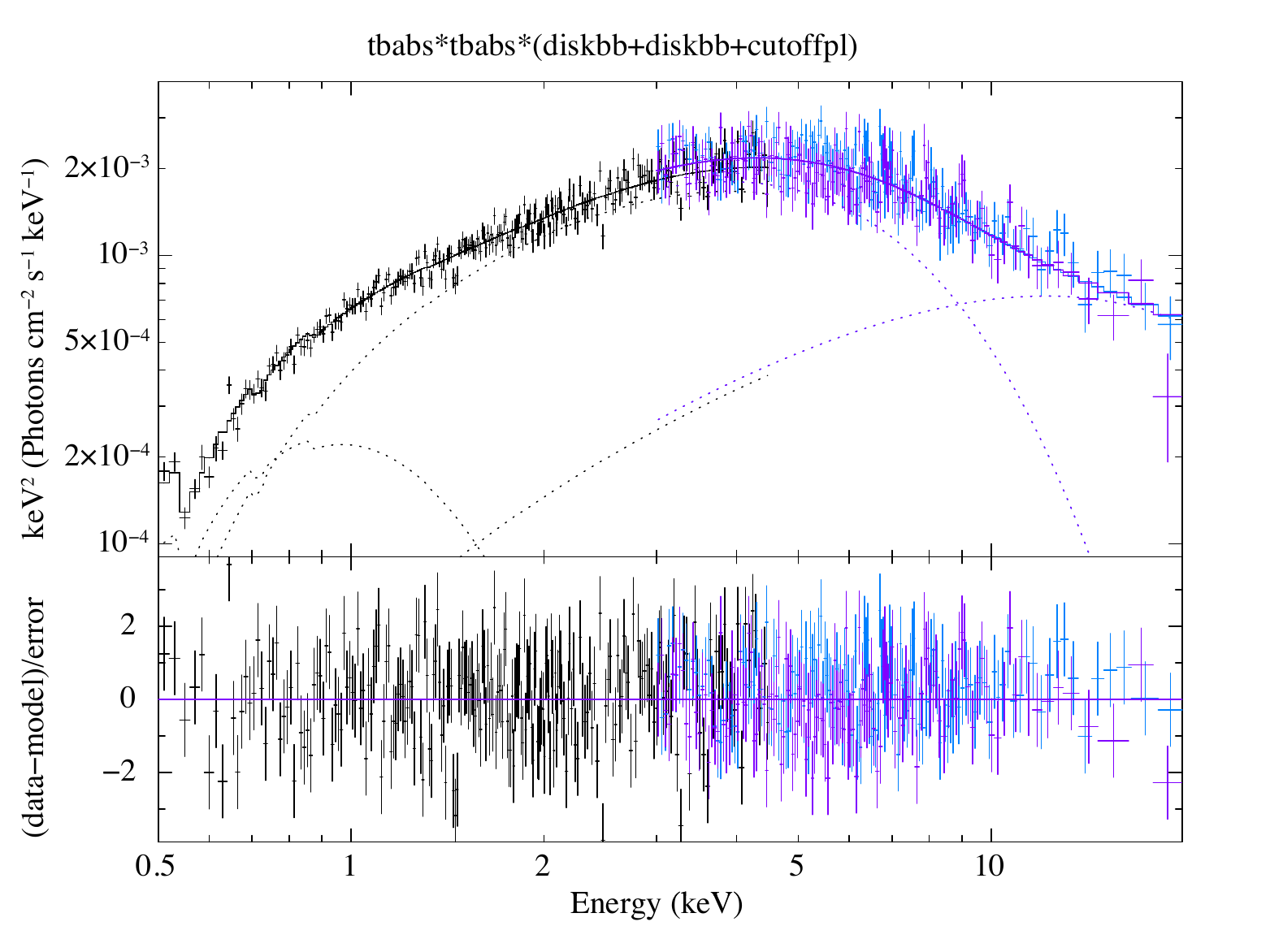}
	\end{center}
	\vspace{-5mm}
	\caption{The unfolded spectrum and residuals for three of the models used to fit the joint \nicer\ (black) and \nustar\ (blue and purple) data. Spectra rebinned for visual clarity. \\
	Top: {\tt tbabs*tbabs*(diskbb+diskpbb)}. A high-energy excess is evident in the residuals. \\
	Middle: {\tt tbabs*tbabs*(diskbb+simpl*diskbb)}. \\
	Bottom: {\tt tbabs*tbabs*(diskbb+diskbb+cutoffpl)}. \label{fig:nnspec}}
\end{figure}

In order to fit this additional curvature, we next used an absorbed two-thermal-component model, {\tt tbabs*tbabs*(diskbb+diskpbb)}, often used to fit `ultraluminous state' ULX spectra, where the hotter broadened disk component is expected to be produced by an inner super-Eddington slim disk, and the cooler disk component originates from a massive outflowing wind, or potentially an outer sub-Eddington thin disk in the case of an intermediate-mass black hole. This offers an improvement on the singular {\tt diskpbb} model, but there is still an obvious excess at energies above 10\,keV, as commonly seen in broadband observations of ULXs (e.g. \citealt{walton14,walton15,mukherjee15}; Fig.~\ref{fig:nnspec}, top). 

To account for the excess high-energy emission, we followed the approach of \citet{walton20} and fitted the spectra with both a magnetic model and a non-magnetic model. The magnetic model, {\tt tbabs*tbabs*(diskbb+diskpbb+cutoffpl)}, assumes the high-energy emission to originate from an accretion column onto a neutron star with a strong magnetic field. Since we do not detect pulsations from this source (see Section~\ref{sec:timing}) and are therefore unable to isolate a pulsed spectrum, we assume typical ULX pulsar parameters $\Gamma = 0.5$ and $E_{\rm cut} = 8.1$\,keV for the cut-off power-law component as in \citet{walton18b}. For the non-magnetic model, {\tt tbabs*tbabs*(diskbb+simpl*diskpbb)}, we assume that the high-energy excess originates instead from Compton up-scattering of disk photons close to the accretor, and so model the component with the {\tt simpl} convolution model \citep{steiner09} applied to the hotter {\tt diskpbb} component.

We find that both the magnetic and non-magnetic models provide similarly good fits to the data, with no further components evident in the residuals. Interestingly, when a high-energy component is added to the model, the {\tt diskpbb} component is no longer required to be broadened, with the radial dependence of the disk temperature $p$ tending towards high values and consistent in each case with a value of $\approx$0.75, and therefore equivalent to a standard {\tt diskbb} model. We therefore also fitted the spectra with models that replace the {\tt diskpbb} components with a second {\tt diskbb} component instead, and find similarly good fits (Fig.~\ref{fig:nnspec}, middle and bottom). Using the {\tt cflux} convolution model, we calculated the 0.3--10\,keV flux to be $5.8\pm0.5\times10^{-12}$\,\fluxcgs, equivalent to a luminosity of $5.8\times10^{39}$\,\lumcgs\ at 2.9~Mpc. This is the value for the {\tt tbabs*tbabs*(diskbb+diskbb+cutoffpl)} model, though the other models incorporating the high-energy excess all give similar fluxes. 

\begin{figure}
	\begin{center}
	\includegraphics[width=7.9cm]{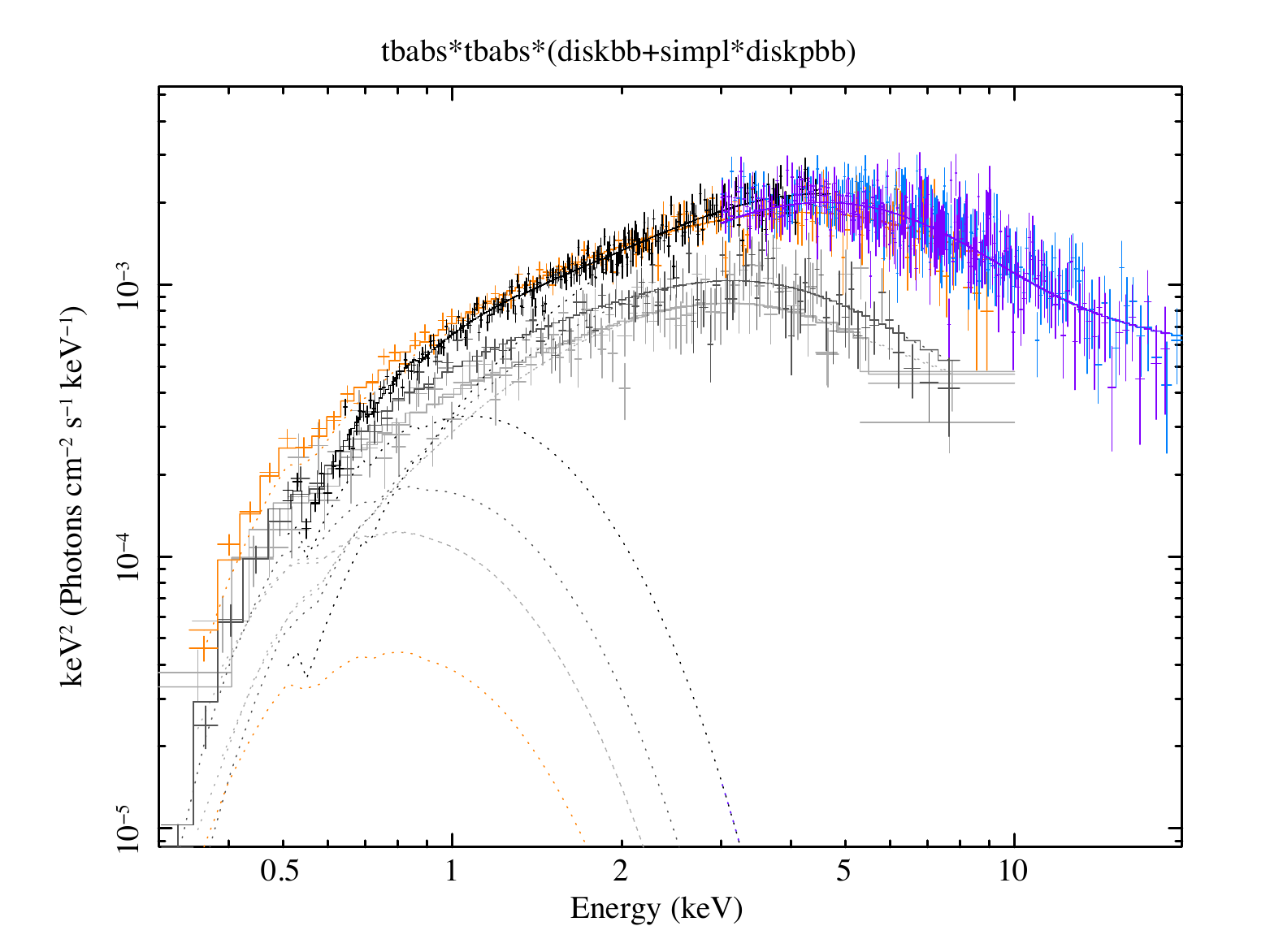}
	\end{center}
	\vspace{-5mm}
	\caption{The unfolded spectra for the three archival \xmm\ observations (0654650101 in light gray, 0654650102 in dark gray, 0654650103 in orange), plotted alongside the \nicer\ and \nustar\ spectra (colors as in Fig.~\ref{fig:nnspec}). \label{fig:xmmspec}}
\end{figure}

\begin{table*}
\caption{The parameters of all multi-component spectral fits to the joint \nicer\ and \nustar\ observation of NGC~4190~ULX-1, as well as for archival \xmm\ observations. Where no uncertainties are given, parameters are frozen.} \label{tab:spectralfitting}
	\vspace{-0.5cm}
	\begin{center}
		\begin{tabular}{@{}l@{}c@{}c@{~}c@{~}c@{~}c@{~}c@{~}c@{~}c@{}c@{~}c@{~}c@{}}
			\hline
			\multicolumn{12}{c}{\nicer\ 3645010101--4 + \nustar\ 30601009002} \\
			Model & $N_{\rm H}^a$  & $T_{\rm in,1}$ & norm$_1$ & $T_{\rm in,2}$ & $p$ & norm$_2$ & $\Gamma$ & $f_{\rm scat}$/$E_{\rm cut}$ & norm$_3$ & $c^b$ & $\chi^2$/d.o.f. \\
			{\tt tb*tb*(...)} & $(\times10^{21}$\,cm$^{-2}$) & (keV) &  & (keV) & & $(\times10^{-3})$ & & --/(keV) & $(\times10^{-5})$ \\
			\hline
			{\tt dbb+dpbb} & $3.9\pm0.7$ & $0.15\pm0.02$ & $400^{+740}_{-290}$ & $2.50\pm0.09$ & $0.60\pm0.02$ & $4.2^{+1.1}_{-0.9}$ & -- & -- & -- & $1.05\pm0.04$ & 1078.5/767 \\
			{\tt dbb+s*dpbb} & $2.4^{+0.4}_{-0.3}$ & $0.28^{+0.04}_{-0.05}$ & $14^{+19}_{-7}$ & $1.2^{+0.3}_{-0.1}$ & $>0.76$ & $210^{+80}_{-150}$ & $3.0^{+0.2}_{-0.3}$ & $0.5\pm0.2$ & -- & $1.23^{+0.06}_{-0.05}$ & 964.7/765 \\
			{\tt dbb+dpbb+cpl} & $2.5^{+0.6}_{-0.5}$ & $0.25^{+0.06}_{-0.05}$ & $19^{+43}_{-12}$ & $1.6\pm0.1$ & $0.76^{+0.15}_{-0.08}$ & $40^{+30}_{-10}$ & $0.5$ & $8.1$ & $7.1\pm0.7$ & $1.09\pm0.04$ & 962.9/766 \\
			{\tt dbb+s*dbb} & $2.7^{+0.5}_{-0.4}$ & $0.23^{+0.03}_{-0.02}$ & $28^{+33}_{-16}$ & $1.5\pm0.1$ & -- & $60^{+20}_{-10}$ & $2.7^{+0.4}_{-0.5}$ & $0.3^{+0.2}_{-0.1}$ & -- & $1.23^{+0.07}_{-0.06}$ & 967.9/766 \\
			{\tt dbb+dbb+cpl} & $2.6\pm0.4$ & $0.25^{+0.03}_{-0.02}$ & $21^{+20}_{-11}$ & $1.61\pm0.06$ & -- & $37^{+5}_{-4}$ & 0.5 & 8.1 & $7.0\pm0.6$ & $1.09\pm0.04$ & 962.9/767 \\
			\hline			
			{\tt dbb+dbb+cpl}$^c$ & $2.6\pm0.4$ & $0.24\pm0.03$ & $26^{+27}_{-14}$ & $1.5\pm0.1$ & -- & $43^{+10}_{-9}$ & $0.4\pm0.5$ & $6^{+4}_{-2}$ & $9.6^{+1.1}_{-0.6}$ & $1.09\pm0.04$ & 965.0/776 \\
			{\tt dbb+dpbb+cpl}$^c$ & $2.3^{+0.5}_{-0.3}$ & $0.29\pm0.05$ & $12^{+21}_{-3}$ & $1.4^{+0.2}_{-0.1}$ & $>0.75$ & $110^{+30}_{-70}$ & $0.3^{+0.4}_{-0.7}$ & $5^{+3}_{-1}$ & $12^{+16}_{-8}$ & $1.09\pm0.04$ & 962.4/775 \\
			\hline
			\multicolumn{12}{c}{\xmm\ 0654650101} \\
			\hline
			{\tt dbb+s*dbb} & 1.0 & $0.20^{+0.09}_{-0.07}$ & $11^{+65}_{-9}$ & $1.10^{+0.09}_{-0.08}$ & -- & $80\pm20$ & 2.7 & 0.3 & -- & -- & 99.2/95 \\
			{\tt dbb+dbb+cpl} & 1.0 & $0.21^{+0.09}_{-0.07}$ & $10^{+50}_{-8}$ & $1.21^{+0.09}_{-0.08}$ & -- & $50^{+20}_{-10}$ & 0.5 & 8.1 & 2.4 & -- & 97.5/95 \\
			\hline
			\multicolumn{12}{c}{\xmm\ 0654650201} \\
			\hline
			{\tt dbb+s*dbb} & $1.5^{+0.6}_{-0.5}$ & $0.22^{+0.07}_{-0.04}$ & $14^{+38}_{-11}$ & $1.11\pm0.05$ & -- & $90\pm20$ & 2.7 & 0.3 & -- & -- & 252.1/246 \\
			{\tt dbb+dbb+cpl} & $1.5^{+0.6}_{-0.5}$ & $0.23^{+0.07}_{-0.04}$ & $12^{+35}_{-10}$ & $1.20\pm0.05$ & -- & $60\pm10$ & 0.5 & 8.1 & 3.7 & -- & 251.8/246 \\
			\hline
			\multicolumn{12}{c}{\xmm\ 0654650301} \\
			\hline
			{\tt dbb+s*dbb} & $0.8\pm0.6$ & $0.5^{+0.2}_{-0.1}$ & $0.8^{+2.1}_{-0.5}$ & $1.6^{+0.2}_{-0.1}$ & -- & $40\pm10$ & 2.7 & 0.3 & -- & -- & 304.7/309	 \\
			{\tt dbb+dbb+cpl} & $0.8^{+0.3}_{-0.2}$ & $0.4^{+0.2}_{-0.1}$ & $0.9^{+2.4}_{-0.6}$ & $1.6\pm0.1$ & -- & $30\pm10$  & 0.5 & 8.1 & 7.5 & -- & 304.9/309 \\
			\hline
		\end{tabular}
	\end{center}
	XSPEC model abbreviations: {\tt tb} = {\tt tbabs}, {\tt cpl} = {\tt cutoffpl}, {\tt dbb} = {\tt diskbb}, {\tt dpbb} = {\tt diskpbb}, {\tt s} = {\tt simpl} \\
	$^a$ $N_{\rm H} $ for the second {\tt tbabs} component; the first is frozen to the Galactic value of $N_{\rm H} = 1.84\times10^{20}$\,cm$^{-2}$. \\
	$^b$ Normalization constant between \nicer\ and \nustar. \\
	$^c$ Cut-off power-law parameters found by fitting simultaneously with the covariance spectrum. 
\end{table*}

While the archival \xmm\ data can be acceptably fitted with single-component models \citep{ghosh21}, we assume that the underlying source spectrum is similar to what we observe for the joint \nicer\ and \nustar\ observation. Therefore we fit the \xmm\ observations with the magnetic and non-magnetic models (using two {\tt diskbb} components to reduce the number of free parameters) in order to trace how the different spectral components change between observations. Since there is no high-energy coverage for these observations, we freeze the parameters of the {\tt simpl} component to $\Gamma=2.7$ and $f_{\rm scat}=0.3$ as in the best-fitting model using two {\tt diskbb} components for the broadband dataset, and freeze the normalization of the {\tt cutoffpl} component to the upper limit found on an initial unfrozen fit so that its presence in the model will have a similar impact on the parameters of the hot thermal component as in the broadband fit. We also freeze the second {\tt tbabs} component to $1\times10^{21}$\,cm$^{-2}$ for the first \xmm\ observation (based on the fits to the other two \xmm\ observations), since there is insufficient low-energy data to achieve good constraints due to the low good exposure time. 

We find that there is no significant difference between the parameters for the thermal components in the magnetic model compared with the non-magnetic model. In both cases, the first two lower-flux observations show lower temperatures in both thermal components than the third observation, which is closer in flux to the broadband epoch. However, the cool thermal component is very similar in temperature between the lower-flux \xmm\ observations and the broadband observation, showing a slight increase in temperature for the third \xmm\ epoch (this may be degenerate with a decrease in $N_{\rm H}$, which we also see for this observation). The hot component temperature is very similar between the third \xmm\ epoch and the broadband spectrum. We plot the \xmm\ spectra alongside the broadband \nicer+\nustar\ spectrum in Fig.~\ref{fig:xmmspec}.

We did attempt to split the data by time based on the slight decrease in flux during the course of the observation (see Fig.~\ref{fig:lc}), although we were not able to find any significant differences in model parameters on this basis. 

\subsection{Timing Analysis} \label{sec:timing}

We generated power spectra using v1.0 of the {\tt stingray} python package for X-ray timing \citep{stingray}. For the \nicer\ observation, we binned the source events within the 0.5--4.5\,keV energy range into a light curve with $dt=0.001$\,s, and broke the light curve into segments of length 524.288\,s ($2^{19}\times dt$). This generated 42 segments, which we used to create an average power spectrum with approximately Gaussian uncertainties, allowing for fitting using $\chi^2$ statistics. The spectrum was geometrically rebinned by a factor of 1.03 and normalized using Leahy normalization (where the Poisson white noise level is set to 2). We followed a similar process for the \nustar\ power spectrum, using the combined FPMA and FPMB events within 3--20\,keV, binning the light curve with $dt=0.001$\,s and creating an average power spectrum using 67 segments of length 1048.576\,s ($2^{20}\times dt$), binned and normalized in the same fashion. Both power spectra were converted to XSPEC format using the {\tt flx2xsp} FTOOL so that they could be fitted using the XSPEC software (see Appendix A1 of \citealt{ingram12}).

\begin{figure}
	\begin{center}
	\includegraphics[width=7.9cm]{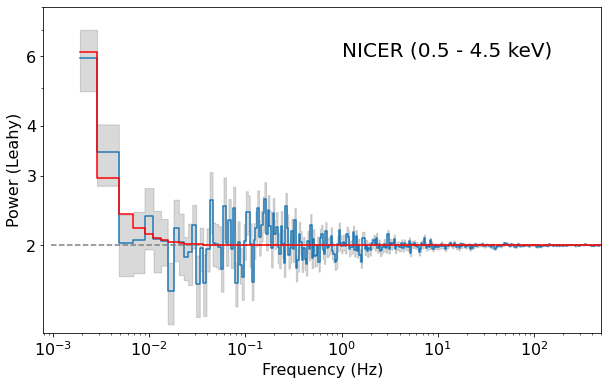}
	\includegraphics[width=7.9cm]{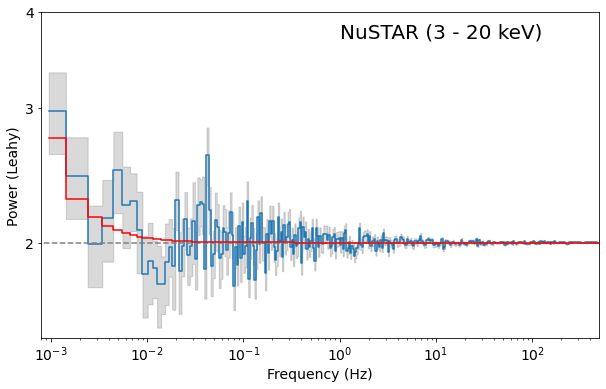}
	\end{center}
	\vspace{-5mm}
	\caption{The power spectra for \nicer\ (top) and \nustar\ (bottom) in Leahy normalization, with errors given in gray filled regions, the Poisson noise level indicated with a dashed gray line, and the best-fitting power-law models plotted in red. \label{fig:powspec}}
\end{figure}

Both power spectra are mostly featureless except for stochastic noise increasing towards lower frequencies below $\sim$0.01\,Hz (Fig.~\ref{fig:powspec}). Using the Whittle statistic \citep{whittle53}, and accounting for the Poisson noise using a flat power-law with normalization set to 2, both can be fitted with a power-law $P(f)\propto f^{-\alpha}$ with $\alpha\approx$1--2, as expected for red noise often seen in accretion processes and seen in some of the ULX population (\citealt{heil09}). For \nicer, we find $\alpha=2.10^{+1.9}_{-0.6}$; for \nustar\ we are unable to obtain strong constraints and simply find $\alpha>0.8$. Neither power spectrum shows any evidence of the presence of a QPO. The fractional rms below 0.1\,Hz is $\sim$0.06 for \nicer\ and $\sim$0.18 for \nustar. In other words, there is a higher contribution to the variability in the \nustar\ band, although there is better signal-to-noise in the \nicer\ band.

\begin{figure}
	\begin{center}
	\includegraphics[width=7.9cm]{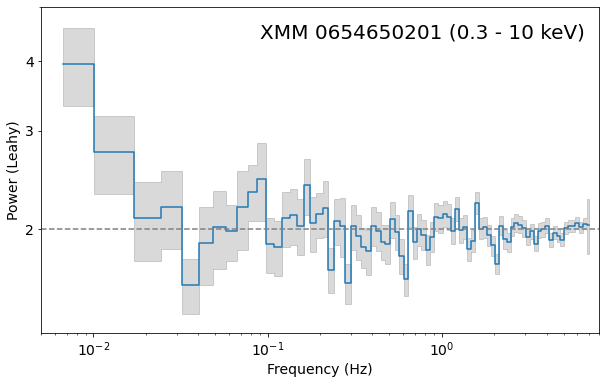}
	\includegraphics[width=7.9cm]{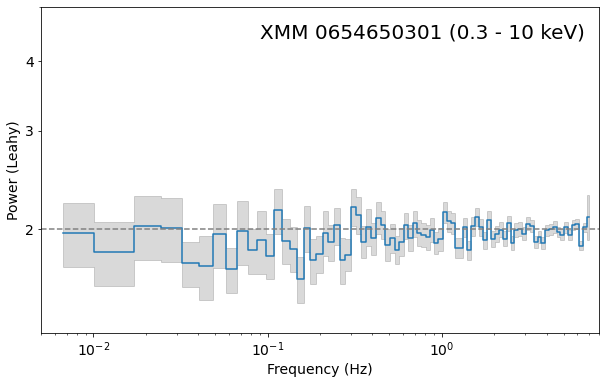}
	\end{center}
	\vspace{-5mm}
	\caption{The power spectra for the \xmm\ observations 0654650201 (top) and 0654650301 (bottom) in Leahy normalization, with errors given in gray filled regions and the Poisson noise level indicated with a dashed gray line. \label{fig:xmm_powspec}}
\end{figure}

We performed a similar analysis for the two archival \xmm\ observations with a significant amount of usable EPIC-pn data (the first observation suffered from large amounts of background flaring and only had 200\,s of good EPIC-pn time), using $dt=0.0734$\,s (i.e. the time resolution of the EPIC-pn camera) and segments of length 150.3232\,s ($2^{11}\times dt$), for 39 and 58 segments for observations 0654650201 and 0654650301 respectively (Fig.~\ref{fig:xmm_powspec}). For 0654650201, we see similar stochastic variability towards low frequencies with $\alpha\approx2$ (although similar to the \nustar\ power spectrum we are unable to place strong constraints on the slope), and the fractional rms below 0.1\,Hz is approximately $\sim$0.17. However, for 0654650301 we do not see any significant variability above the Poisson noise level. 

While there are no obvious pulsation signals in the power spectra, processes such as spin-up or orbital modulation often mean that genuine pulsations do not show up in a straightforward power spectrum. Therefore we ran an accelerated pulsation search on the \nicer, \nustar, and the latter two \xmm\ observations using the {\tt HENaccelsearch} routine from v7.0 of the HENDRICS timing software \citep{hendrics}, searching for pulsations between 0.01 and 10\,Hz with acceleration between 10$^{-10}$ and $5\times10^{-9}$ Hz\,s$^{-1}$. We did not find any significant signals in any of the observations using this method.

The reasonably high number of photons from the \nicer\ and \nustar\ observations allows us to place relatively constraining upper limits on the pulsed fraction. To estimate an upper limit on the pulsed fraction, we used {\tt stingray} to simulate light curves matching the good time intervals and average count rate of both observations, with the addition of a sinusoidal pulsation signal at a period of 1\,s fairly typical of ULX pulsars (e.g. \citealt{bachetti14,israel17,sathyaprakash19}) at a variety of pulsed fractions. We simulated 100 light curves for each pulsed fraction in the range $10\% < PF < 50\%$ in steps of 0.5\% and searched for pulsations using epoch folding (we did not simulate any kind of acceleration as this ultimately has no effect on the upper limit, and allows us to use a more straightforward pulsation search). We define the upper limit as the pulsed fraction for which at least 90\% of simulations detected the pulsation with a significance of at least 3$\sigma$. We find an upper limit on the pulsed fraction of 16\% for \nicer\ (0.5 -- 4.5\,keV) and 35\% for \nustar\ (3 -- 20\,keV).

\begin{figure}
	\begin{center}
	\includegraphics[width=7.9cm]{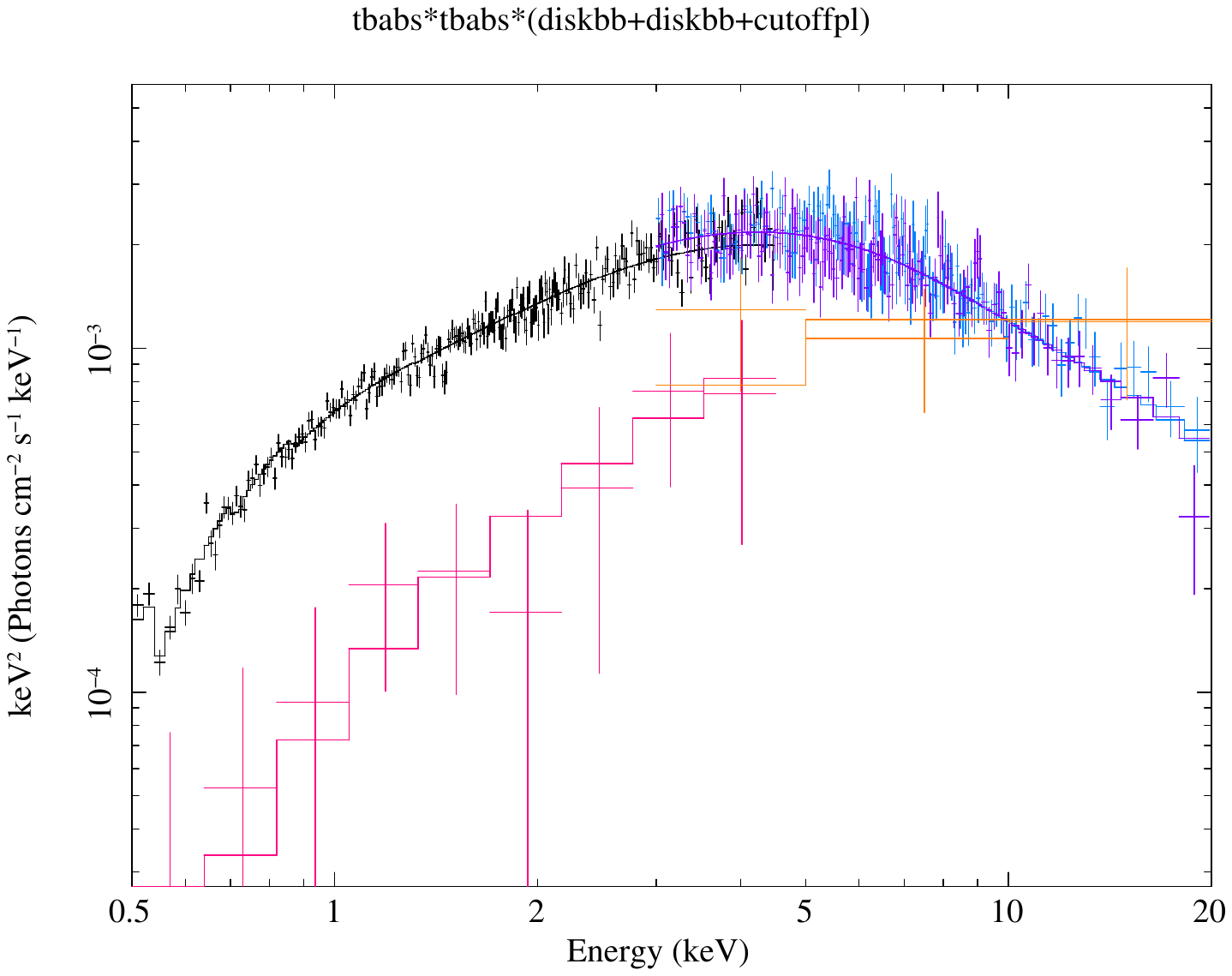}
	\end{center}
	\vspace{-5mm}
	\caption{The \nicer\ and \nustar\ energy spectrum (colors as in Fig.~\ref{fig:nnspec}), plotted with the covariance spectrum (magenta and orange for \nicer\ and \nustar\ respectively).  \label{fig:covar}}
\end{figure}

To further explore the variability of NGC~4190~ULX\=/1, we computed the covariance spectrum \citep{wilkinson09}, shown in Fig.~\ref{fig:covar}. The covariance spectrum shows correlated variability as a function of energy, which may allow us to identify the spectral components that are the source of coherent variability, and has previously been used to disambiguate the multiple components of the energy spectra of ULXs \citep{middleton15,middleton19,kara20}. We find that the covariance increases towards higher energies, although it does not appear to directly match the spectral shape of the hot thermal component -- if fitted with a {\tt diskbb} or {\tt diskpbb} model we find that it requires temperatures of $T_{\rm in} > 3$\,keV, far higher than that found for $T_{\rm in,2}$ regardless of the model used to fit the spectrum. The covariance spectrum does, however, appear to resemble the cut-off power-law component used to model a potential accretion column in the magnetic model, although by itself it cannot constrain a cut-off energy, so we proceeded to fit it simultaneously with the broadband energy spectrum, including an additional normalization constant to account for the lower normalization expected for the covariance spectrum, and tying the parameters of the {\tt cutoffpl} component to those fitted to the covariance spectrum.

We find that, with the {\tt cutoffpl} parameters so constrained, the lower-energy portion of the spectrum still requires two thermal components to be fitted well. Attempting to fit the spectrum with the high-energy cut-off power-law and a single softer thermal component (whether broadened or not), or with a {\tt diskbb+nthcomp} model similar to that used by \citet{middleton15}, results in a poorer fit ($\chi^2>1000$ in each case) and the `m-shaped' residuals in the \nicer\ data indicating the presence of two components. Therefore, using the {\tt tbabs*tbabs*(diskbb+diskbb+cutoffpl)} model, we find the cut-off power-law parameters $\Gamma = 0.4\pm0.5$ and $E_{\rm cut} = 6^{+4}_{-2}$\,keV, very similar to those earlier assumed to be representative of a ULX pulsar accretion column. The parameters for the thermal components were also found to be similar to the previous broadband spectral fit. We also tried fitting the hotter of the two thermal components with a {\tt diskpbb} model to see whether there is any evidence of broadening when the cut-off power-law parameters are constrained by the covariance spectrum. In this case as well as before, the radial temperature dependence $p$ tends towards the maximal value of 1 (indicating a narrower spectral component) with a lower limit of $p > 0.75$, ruling out a spectrally broadened disk component.

\section{Discussion} \label{sec:disc}

NGC~4190~ULX-1 has a spectrum typical of the overall ULX population, with two thermal components below 10\,keV and an additional hard excess. Its spectrum is generally hard, with the hotter thermal component the dominant of the two, and there are no significant residuals around 1\,keV as found in some ULXs more dominated by soft emission \citep{middleton14,middleton15b}, which are indicative of powerful, relativistic outflowing winds \citep{pinto16}. This suggests that we are viewing the source at a relatively low inclination, with most of the emission originating from accretion at the center of the system. There is a clear high-energy excess above 10\,keV, although we cannot determine whether the origin of this component is magnetic or non-magnetic from the spectrum alone, as both models provide fits of similar quality. The particularly low-energy turnover for a ULX appears to result from the hot thermal component not showing the broadening typically exhibited in ULX spectra, with values of $p$ consistent with or higher than the standard thin-disk value of $p=0.75$. 

Similar to other ULXs that show significant stochastic variability (which is not a universal feature of ULXs, found in less than half of the population; \citealt{heil09}), NGC~4190~ULX-1 demonstrates red noise below 0.1\,Hz in multiple observations. Many of the ULXs found to have high levels of variability have soft spectra (e.g. \citealt{sutton13}), though the covariance spectra of such sources \citep{middleton15} indicate that the variability is associated with the hotter thermal component, suggesting a model in which variability is imprinted upon the hard emission by a clumpy outflowing wind along the line of sight \citep{middleton11,takeuchi13}. 

In contrast to these soft variable ULXs, NGC~4190 ULX\=/1 has a hard spectrum, and the covariance spectrum shows that its variability is not associated with the hot thermal component but is associated with the harder excess emission, and appears to confirm its cut-off power-law nature. NGC~4190~ULX\=/1 is not unique in being a hard variable ULX -- for example, M51~ULX-7 also has a hard spectrum and significant short-term variability, showing flicker noise as well as a low-frequency break in its power spectrum below $10^{-3}$\,Hz \citep{earnshaw16}, and was later found to be a ULX pulsar \citep{rodriguezcastillo19}. Another example in the same galaxy is M51 ULX-8, a reasonably hard and variable source which is not a pulsating ULX but still proposed to be a neutron star accretor from the detection of a potential cyclotron absorption line in its spectrum \citep{brightman18}, and whose covariance spectrum shows that its variability also originates in a hard cut-off power-law component consistent with an accretion column, and possibly resulting from a combination of accretion column and inner disk emission \citep{middleton19}. 

The covariance spectrum of NGC~4190~ULX-1 suggests that the variability in this source originates not from its hot thermal component but from a cut-off power-law component with parameters similar to those typically found in ULX pulsars (e.g. \citealt{walton18b}). Therefore, despite the lack of detected pulsations, the timing properties of NGC~4190~ULX-1 nevertheless potentially provide indirect evidence of the presence of an accretion column component. It is possible for a high-energy power-law excess to be generated by an accreting black hole ULX in the absence of an accretion column (e.g. \citealt{mills23}), and indeed \citet{combi24} find the spectral energy distribution of NGC~4190~ULX-1 to be consistent with a model that assumes a 10\,M$_\odot$ black hole accreting at 10\,$\dot{M}_{\rm Edd}$, but in this case it is not clear whether the covariance spectrum would take the cut-off power-law form that we see here. If the accretor is indeed a neutron star with an accretion column, it is possible that the magnetic field and spin alignment of the neutron star are not favorable to produce pulsations. 

We also note that the lack of detected pulsations do not rule out the presence of pulsations altogether. While we can rule out extremely high pulsed fractions such as the pulsed fraction of 72\% found for NGC~300~ULX\=/1 in the \nustar\ band \citep{carpano18} compared with our upper limit of 35\% for NGC~4190~ULX-1 in the same band, several ULX pulsars have lower pulsed fractions than this even where they peak in the \nustar\ band. For example, pulsations of NGC~7793~P13 have been found to peak in the \nustar\ band at pulsed fractions of $\sim$20--60\% at different times \citep{fuerst16,fuerst21}, and in the \xmm\ band pulsations have been detected at pulsed fractions as low as a few percent (e.g. in NGC~1313~X-2; \citealt{sathyaprakash19}), lower than our \nicer\ upper limit of 16\%. Additionally, pulsed fractions lower even than this may well be present in the ULX population and simply undetected due to the lack of sufficient data to detect them, given that detecting pulsations can be challenging anyway due to spin-up and orbital effects. (We note that \citealt{combi24}, following a different method, found pulsed fraction upper limits of 7\% and 18\% for \nicer\ and \nustar\ respectively at the 90\% significance level, which are approximately equivalent to our 3$\sigma$ upper limits.)

The matter is further complicated by the fact that pulsations in ULX pulsars are sometimes found to be transient (e.g. \citealt{bachetti20}), such that multiple high-quality observations may be required to successfully identify a genuine ULX pulsar. Future observations with \nustar\ in particular, with a moderate (50--100\%) increase in exposure time, would be able to probe pulsed fractions down to 30\% at energies above 10\,keV (the energy range in which they are strongest in ULX pulsars), as well as provide more opportunities to search for transient pulsations.

Given the low frequency of the aperiodic variability we detect, it is unlikely to be intrinsic to an accretion column, which has a much shorter characteristic timescale than these observations are sensitive to, and variability is at any rate expected to be suppressed within the magnetospheric radius \citep{revnivtsev09, mushtukov19}. Instead, the variability could be imprinted upon the high-energy emission by a clumpy obscuring wind as described in \citet{middleton15}. This is in tension with the apparent low inclination of this source we mention at the beginning of this section, but given that the presence of a soft thermal component is evident, the inclination may be high enough for a small amount of the outflowing wind to intercept the line of sight to the center of the system and to make a modest contribution to the energy spectrum. At a favorable inclination, it may also be possible for most of the hot thermal emission from the super-Eddington disk to remain unobscured, leading to the component not making a significant contribution to the covariance spectrum. 

The disappearance of short-term variability in the third \xmm\ observation appears to be connected with an increase in the temperature of the soft thermal component and/or a decrease in the $N_{\rm H}$ of the absorption. In the case that the variability originates from the high-energy emission being partially obscured by a clumpy wind, this may indicate a decrease in the spherization radius $R_{\rm sph}$ due to a lowered mass accretion rate, leading to the clumpy wind no longer crossing the line of sight to the observer. The lack of high-energy data during the third \xmm\ observation makes it difficult to determine the true contribution of a cut-off power-law component -- while its normalization upper limit is consistent with what we measure for the broadband observation, we cannot confirm its presence from the \xmm\ data alone, and it is possible that the high-energy component is itself absent in this observation. Further broadband observations of the source in multiple flux states would provide further insight into the behavior of the hard emission over time. 

The long-term variability in the spectrum appears to be driven primarily by the temperature and flux of the hot thermal component rather than the cool component. Also, while we only have \nustar\ data for one of the epochs, the low-energy spectral turnover makes it clear that this ULX does not have a very stable high-energy tail (unlike that seen in some ULXs such as Ho~IX~X-1 and NGC~1313~X-1, which show considerable soft variability yet very consistent high-energy tails; \citealt{walton17,walton20}). For a non-magnetic model in which the high-energy emission originates from up-scattering of photons from the hot thermal component, this variability arises naturally from the changes in disk temperature changing the intrinsic spectrum that is up-scattered. For the magnetic model, it implies a change in the flux of the accretion column component, potentially due to a changing accretion column height resulting from changes in the mass accretion rate through the column, or changing obscuration of the accretion column. 

\begin{figure}
	\begin{center}
	\includegraphics[width=7.9cm]{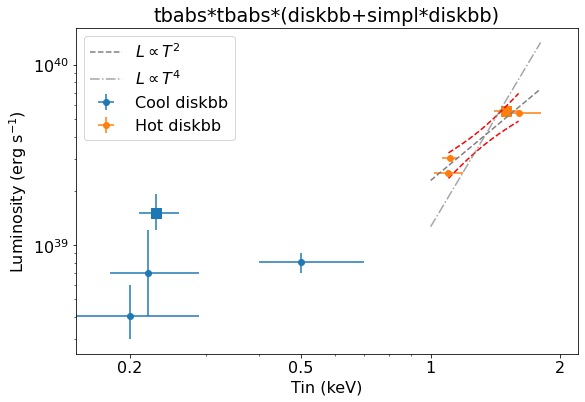}
	\includegraphics[width=7.9cm]{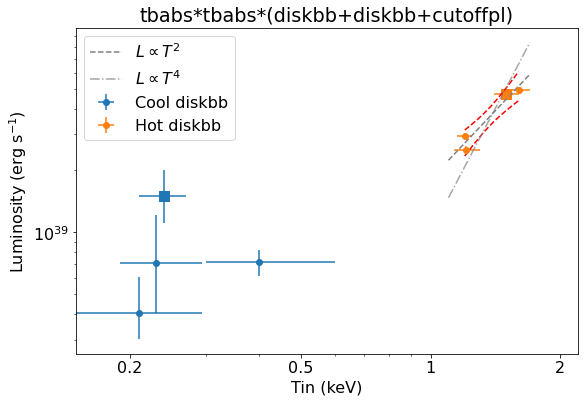}
	\end{center}
	\vspace{-5mm}
	\caption{Unabsorbed luminosity-temperature relations for NGC 4190 ULX-1 when fitted with the non-magnetic (top) and magnetic (bottom) models. The cool thermal component is plotted in blue, and the hot component in orange. The more reliable values obtained from the broadband observation are indicated with squares. The dark grey and red dashed lines indicating the line of best fit to the hot component and its errors respectively, with a $L\propto T^4$ relation plotted with a lighter grey dot-dashed line for comparison. \label{fig:lt}}
\end{figure}

We can explore the spectral variability further by making luminosity-temperature plots for the two thermal components (Figure~\ref{fig:lt}). The XSPEC model {\tt cflux} evaluated between 0.001 and 100\,keV was used to estimate the unabsorbed bolometric luminosity for the two {\tt diskbb} components in the {\tt diskbb}-only fits (these chosen to reduce the number of free parameters). The constraints placed using the covariance spectrum fit were considered in the case of the broadband \nicer\ and \nustar\ observation. No obvious trend is seen for the cool thermal component save for the higher-temperature observation appearing to be a distinct outlier. We note that the luminosities seen for this component are super-Eddington or nearly so, particularly in the case of a neutron star accretor, which suggests that this component results from a soft outflowing wind, ruling out a thin, sub-Eddington outer accretion disk for this source.

In contrast, the hot thermal component appears to demonstrate a clear relationship between luminosity and temperature, which we found to be $L\propto T^{2.0\pm0.9}$ via least-squares fitting for the non-magnetic model, and similarly $L\propto T^{2.2\pm1.1}$ for the magnetic model. This result comes with obvious caveats -- namely that the \xmm\ spectra are overfitted by three-component models and are unable to constrain any of the high-energy component parameters, therefore the uncertainties are underestimated and some parameters potentially biased due to others being frozen. We are also working with a small number of observations clustered together in $L$-$T$ space, insufficient to make strong claims about the behavior over a wide range of luminosities and disk temperatures. A higher number of truly broadband observations would be required to confirm this relationship. 

However, if we take the relation at face value, it is interesting to note that despite there being no evidence for the spectral broadening expected from a slim, super-Eddington accretion disk, the $L$-$T$ relation is more  similar to the shallow $L\propto T^2$ relation expected for a slim disk \citet{watarai00}, compared with the $L\propto T^4$ expected for a standard sub-Eddington accretion disk \citet{shakura73}, implying that a slim disk may still be present (\citealt{combi24} also favor a shallow $L\propto T^2$ relation over a $L\propto T^4$ relation in their analysis). It is possible that such a spectrally narrow thermal component could originate from a slim disk if it extends only over a small number of radii. We would normally expect a slim disk to reach all the way to the innermost stable circular orbit (ISCO) at the high mass accretion rates that would produce it, but the presence of a strong enough magnetic field from a neutron star accretor may truncate the disk before it reaches the ISCO (as suggested by \citealt{walton18a}), which would result in a narrower range of temperatures and an artificially high value of $p$ when fitting with a {\tt diskpbb} model. Therefore, if present, this shallow $L$-$T$ relation combined with a narrow hot component may be tentative further evidence of the presence of a neutron star accretor, with a magnetospheric radius close to but slightly less than the spherization radius of the super-Eddington accretion disk. Ultimately, further broadband data is required to confirm this behavior.

\section{Conclusions} \label{sec:conc}

We present the spectral and timing analysis of a broadband \nicer+\nustar\ observation of the nearby and relatively under-studied NGC~4190~ULX-1, which exhibits spectral features typical of the wider ULX population and significant spectral variability. While pulsations were not detected, the presence of a hard cut-off power-law component in the covariance spectrum and the luminosity-temperature properties of the narrow hot thermal component provide tentative evidence towards a neutron star accretor with an accretion column and a super-Eddington accretion disk truncated by the neutron star magnetic field. 

The compelling possibility of NGC~4190~ULX-1 hosting a neutron star accretor, coupled with its nearby and highly accessible nature, make it an ideal target for future ULX study. Further broadband observations will be critical in confirming the luminosity-temperature trends suggested by the archival \xmm\ observations, for investigating the connection between the changing spectral and timing properties of this source, and for further exploring the nature and behavior of its high-energy emission. Future observations will also be necessary to continue the search for pulsations, in the case that they are simply transient in this source, or other properties such as cyclotron lines that may be able to confirm the neutron star nature of this source \citep{brightman18,walton18c}.

While \nicer\ is not typically used for ULX studies due to their comparatively low fluxes compared with the \nicer\ background and their extragalactic nature making source confusion more likely, it is clear that it can still be a powerful tool for the timing analysis of the few ULXs that are nearby and reasonably well isolated. However, a full picture of the X-ray behavior of this source can only be obtained with broadband X-ray data. While coordination of a soft X-ray observatory with \nustar\ is invaluable with present-day capabilities, this source illustrates the value of a future dedicated broadband observatory such as the \textit{High Energy X-ray Probe} (\textit{HEX-P}) in the study of ULXs \citep{madsen23,bachetti23}. 

\acknowledgments

We thank the anonymous referee for useful comments to improve the clarity of this article. HPE acknowledges support under NASA grant 80NSSC21K0123 and NASA contract NNG08FD60C. The majority of this work was performed on the traditional homeland of the Tongva people.

This work made use of data from the \nustar\ and \nicer\ missions, funded by the National Aeronautics and Space Administration. We thank the observing teams for carrying out the coordinated observations. This work has also made use of archival observations by \xmm, an ESA science mission with instruments and contributions directly funded by ESA Member States and NASA, and the \textit{Neil Gehrels Swift Observatory}, a NASA science mission. 

%

\vspace{5mm}
\facilities{NICER, NuSTAR, XMM, CXO, HEASARC, Swift(XRT)}
\software{astropy \citep{astropy13,astropy18}, HENDRICS \citep{hendrics}, HEASoft \citep{heasarc14}, NICER software, NuSTARDAS, \xmm\ SAS, Stingray \citep{stingray}} 

\vspace{1cm}
\bibliography{ngc4190ulxpaper}
\bibliographystyle{../aasjournal}

\end{document}